\documentstyle{mn}

\title{Mode Identification of the Slowly Pulsating F0V Star
  V398 Aurigae (9 Aur)}
\author[C. Aerts and K. Krisciunas]
       {C. Aerts$^{1,}$\thanks{Senior Research Assistant, Belgian National
Fund for Scientific Research}, K. Krisciunas$^2$\\ $^1$Instituut voor
Sterrenkunde, Katholieke Universiteit Leuven, Celestijnenlaan 200 B,
B-3001 Heverlee, Belgium \\ $^2$Joint Astronomy Centre, 660 N.
A`oh\={o}k\={u} Place, University Park, Hilo, HI 96720, USA}

\date{Accepted 1995 September 15.
      Received 1995 MMM nn; in original form 1995 May 15}

\begin{document}

\maketitle

\begin {abstract}
We have investigated the F0V star V398\,Aurigae (= 9 Aur) under the assumption
that it is undergoing non-radial gravity mode oscillations and that the two
principal periods given by Krisciunas et al. (1995) are correct.  We find that
the two periods are manifestations of an $\ell=3, |m|=1$ spheroidal mode and
its toroidal corrections due to the rotation of the star.  As far as we know,
this is the first detection of toroidal correction terms in a real star.  The
two modes probably are the result of rotational splitting.

Our analysis provides for the first time a physical explanation of certain
characteristics of the observed behavior of the star.
The amplitude of the radial part of the pulsation
for $f_1 = 0.795$\,d$^{-1}$ is a factor of 4 larger than the one for $f_2 =
0.346$\,d$^{-1}$.  Since the photometric variability is determined mostly by
temperature variations, which in turn are determined by the radial part of the
pulsation, the photometric variability is dominated by the mode with frequency
$f_1$.  On the other hand, $f_2$ is the more pronounced one in all three
spectroscopic moment variations (including the radial velocity), reflecting
that the transverse displacement of $f_2$, and not the one of $f_1$, dominates
the velocity behavior.
\end{abstract}

\section{Introduction}
Krisciunas \& Handler (1995) have compiled a list of 17 stars of similar
spectral type and luminosity class which appear to constitute a new class
of pulsating variable stars.  These stars are typically of spectral type
F0 to F2, and they are found on, or just above, the main sequence in the
Hertzsprung-Russell Diagram.  The best studied examples are $\gamma$
Doradus (Balona, Krisciunas \& Cousins 1994), 9 Aurigae (Krisciunas et al.
1995), HD 224638 and HD 224945 (Mantegazza, Poretti \& Zerbi 1994).  Given
that $\gamma$ Doradus is the brightest member of the list, and that it was
the first one found to be variable (Cousins \& Warren 1963), it has been
suggested that these stars be known as ``$\gamma$ Doradus stars''.

The photometric time scales of the $\gamma$ Dor stars are 0.5 to 3.5 d --
variations an order of magnitude slower than the fundamental radial
pulsation period for stars of this density.  9 Aur and $\gamma$ Dor have
shown evidence for radial velocity and spectroscopic line profile
variations.  In the case of 9 Aur  -- recently designated V398\,Aur in the
latest named list of new variable stars (Kazovarets \& Samus 1995) -- the
radial velocity variations appear to follow only one of the two
photometric frequencies ($f_{2}$ = 0.346 d$^{-1}$) found by Krisciunas et
al. (1995).

Given the time scale of variations, the evidence for radial velocity and
line profile variations, and the lack of other viable explanations
(including star spots), the evidence is strong that $\gamma$ Dor stars are
exhibiting non-radial gravity mode pulsations.  This has been a surprising
observational development, because until now only much hotter stars have
shown evidence for non-radial $g$-modes (Waelkens 1991).

Because the power spectrum of the radial velocities of V398\,Aur did not
show evidence for the highest peak in the power spectrum of the photometry
($f_{1}$ = 0.795 d$^{-1}$), Krisciunas et al. (1995) suggest that $f_{2}$
arises from an $\ell$ = 1 or 2 spherical harmonic, while $f_{1}$ arises
from a higher degree harmonic.  In this paper we apply the moment method
described by Aerts et al.\ (1992) to the line profiles of V398\,Aur in an
attempt to identify its pulsational mode(s).

\section{Cross-correlation profiles}
\label{crosscorr}
Krisciunas et al. (1995) present 95 cross-correlation profiles obtained by
R. F. Griffin with the Haute Provence Coravel (Baranne, Mayor \& Poncet
1979).  While the Coravel is used primarily for the determination of the
radial velocities of stars, the cross-correlation profiles give us,
effectively, a measure of the mean spectroscopic line depths and a
parameter characterizing the width of the lines.  The width parameter can
be interpreted in terms of the projected rotational velocity $v$ sin $i$
of the star (Benz \& Mayor 1981).  The cross-correlation profiles can be
thought of as the mean line profiles of the star being studied.

The Coravel data reduction package also gives us a Gaussian fit to the
points of the cross-correlation profiles.  In order to carry out the
identification of the pulsational mode(s) of a star, it is necessary to
have a very high signal to noise ratio in spectroscopic parameter space.
We found it necessary to use the mathematical {\em fits} to the
cross-correlation points for our analysis here.  (See the sets of curved
lines in Fig 10 of Krisciunas et al. 1995.)

\section{Determination of pulsational mode(s)}
Mode identification is currently often obtained by means of spectroscopic
analyses.  The moment method used here was first introduced by Balona
(1986) and was subsequently generalized by Aerts et al.\ (1992) and by
Mathias et al.\ (1994) in the case of respectively a mono- and a multiperiodic
pulsation. It is based on the time variations of the first three moments
of a line profile.  The periodograms of the three moments can immediately
be interpreted in terms of the periods and amplitudes of the non-radial
pulsation (NRP) parameters.  The observed moment variations are compared
with theoretically calculated expressions for these variations in the case of
various pulsation modes. Mode identification is obtained by means of a
so-called discriminant, which captures the discrepancy between the
observed and the calculated amplitudes.

The discriminant used in this paper is a generalization of the one
proposed by Aerts et al.\ (1992) in the sense that it takes into account
all seven observed amplitudes of the three moments (instead of only three)
and it gives them a weight according to their uncertainty.  This
discriminant turns out to be more accurate than the one presented by Aerts
et al., especially in the case of non-axisymmetric modes (Aerts, in
preparation).  Identification is achieved by means of the minima of the
discriminant, which, in this paper, are defined by~:
\begin{equation}
\label{discr}
\gamma_{\ell}^m\equiv\min_{v_{\rm p},i,\sigma}
\Gamma_{\ell}^m(v_{\rm p},i,\sigma),
\end{equation}
where $v_{\rm p}$ is proportional to the pulsation velocity amplitude (the
proportionality constant is given in Aerts et al.\ 1992), $i$ is the
inclination angle, and $\sigma$ is the width of the intrinsic profile that
is assumed to be Gaussian.  The best solution in the space of the
pulsation parameters $(\ell,m,v_{\rm p},i,\sigma)$ is the one with the
lowest minimum $\gamma_{\ell}^m$. We recall that the discriminant is
unable to find the sign of the azimuthal number $m$, and that it is an
accurate identification tool for low-degree modes ($\ell\leq 4$).  In
constructing the discriminant given by equation\,(\ref{discr}), we have
assumed that the $v\sin i$-value of V398\,Aur is accurately known and
adopt 17.8\,km s$^{-1}$ (Krisciunas et al.\ 1995).  However, since these
authors claim that the non-radial pulsations of the star are giving rise
to line width variations, they point out that the ``true'' projected
rotational velocity may actually be smaller than the mean observed mean
value of 17.8\,km s$^{-1}$, perhaps as small as the smallest observed
values of the line width ($\approx$ 15 km s$^{-1}$).

\begin{figure*}
\vspace{15truecm}
\caption[]{The first, second, and third moment of the mathematical fits to the
cross-correlation profiles of V398\,Aur taken between 1993 December 25 and
1994 January 10 UT (see also in Section\,\ref{crosscorr}). The dots represent
the observations, while the full lines are the fits for a model with
$f_2$ (left) and $f_1$ (right).}
\label{mom}
\end{figure*}

\begin{table*}
\caption{The different minima of the discriminants for the two modes of V398
Aur.  $\gamma_{\ell}^m$, $v_{\rm p}$, and $\sigma$ are given in km
s$^{-1}$.  The radial component of the velocity has an amplitude
proportional to $v_{\rm p}$, while the transverse component's amplitude is
proportional to $Kv_{\rm p}$ (see also Discussion).}
\begin{center}
\begin{tabular}{|c|c|c|c|c|c||c|c|c|c|c|c|}
\hline
\multicolumn{6}{|c||}{} &
\multicolumn{6}{|c|}{} \\[-10pt]
\multicolumn{6}{|c||}{$f_1=0.795$\,d$^{-1}$}&
\multicolumn{6}{c|}{$f_2=0.346$\,d$^{-1}$}\\[2pt]
\hline
\hline
\multicolumn{1}{|c|}{}&
\multicolumn{1}{c|}{}&
\multicolumn{1}{c|}{}&
\multicolumn{1}{c|}{}&
\multicolumn{1}{c|}{}&
\multicolumn{1}{c||}{}&
\multicolumn{1}{c|}{}&
\multicolumn{1}{c|}{}&
\multicolumn{1}{c|}{}&
\multicolumn{1}{c|}{}&
\multicolumn{1}{c|}{}&
\multicolumn{1}{c|}{}\\[-10pt]
\multicolumn{1}{|c|}{$\ell_1$}&
\multicolumn{1}{c|}{$|m_1|$}&
\multicolumn{1}{c|}{$\gamma_{\ell_1}^{m_1}$}&
\multicolumn{1}{c|}{$v_{\rm p}^1$}&
\multicolumn{1}{c|}{$i$}&
\multicolumn{1}{c||}{$\sigma$}&
\multicolumn{1}{c|}{$\ell_2$}&
\multicolumn{1}{c|}{$|m_2|$}&
\multicolumn{1}{c|}{$\gamma_{\ell_2}^{m_2}$}&
\multicolumn{1}{c|}{$v_{\rm p}^2$}&
\multicolumn{1}{c|}{$i$}&
\multicolumn{1}{c|}{$\sigma$}\\
\multicolumn{1}{|c|}{}&
\multicolumn{1}{c|}{}&
\multicolumn{1}{c|}{}&
\multicolumn{1}{c|}{}&
\multicolumn{1}{c|}{}&
\multicolumn{1}{c||}{}&
\multicolumn{1}{c|}{}&
\multicolumn{1}{c|}{}&
\multicolumn{1}{c|}{}&
\multicolumn{1}{c|}{}&
\multicolumn{1}{c|}{}&
\multicolumn{1}{c|}{}\\[-10pt]
\hline
&&&&&&&&&&&\\[-10pt]
3&1&0.92&0.097&$55^{\circ}$&10.1&3&1&1.23&0.023&$56^{\circ}$&10.0\\
4&1&0.93&0.091&$32^{\circ}$&8.4&4&1&1.31&0.025&$66^{\circ}$&5.0\\
3&2&0.97&0.108&$82^{\circ}$&10.1&1&1&1.32&0.030&$29^{\circ}$&4.9\\
2&1&1.01&0.093&$83^{\circ}$&10.6&2&1&1.36&0.008&$56^{\circ}$&8.2\\
2&2&1.02&0.080&$33^{\circ}$&4.8&2&2&1.40&0.019&$32^{\circ}$&4.9\\
$\vdots$&$\vdots$&$\vdots$&$\vdots$&$\vdots$&$\vdots$&
$\vdots$&$\vdots$&$\vdots$&$\vdots$&$\vdots$&$\vdots$\\[2pt]
\hline
\end{tabular}
\end{center}
\label{gamma}
\end{table*}
We here apply the biperiodic moment method to the profiles of V398\,Aur.
The three observed moment variations are shown as dots in Fig\,\ref{mom},
where they are put in phase according to $f_1$ (right) and $f_2$ (left)
respectively.
The peak-to-peak amplitude of the radial velocity (=first moment) amounts
to 3.0\,km s$^{-1}$ for $f_1$ and to 5.2\,km s$^{-1}$ for $f_2$.
It is clear that a frequency analysis of the radial-velocity data cannot
yield the same accuracy as the photometry presented in Krisciunas et al.\
(1995).  Nevertheless, we performed a period analysis on the three
moments of V398\,Aur to check if $f_1$ and $f_2$ are present in them.
We used the Phase Dispersion Minimization method of Stellingwerf (1978)
and found that the frequency $f_2$ is present in all three moment variations,
especially in the odd ones. The presence of the frequency $f_1$ is less
clear, but beat- and sum-frequencies of $f_1$ and $f_2$ are present in the
second and third moment. We further studied the variations of the moments
assuming that the two frequencies $f_1=0.795\,$d$^{-1}$ and
$f_2=0.346\,$d$^{-1}$ are accurate. The solid lines shown in
Fig\,\ref{mom} are the theoretical fits to the observations for a model
with $f_1$ (right) and $f_2$ (left).
The fact that the third moment is centered
around 0\,km s$^{-1}$ reflects that there is no variability on a time
scale larger than the time span of our data.

We can determine the ratio of the horizontal to the vertical pulsation velocity
amplitude $K$ by means of the boundary conditions as a function of stellar mass
and radius if we know the pulsation frequency in the corotating frame.  Since
we only know the frequencies in the observer's inertial frame, we use these to
approximate the true $K$-values.  The actual mass of 9 Aur A, the primary, is
not known.  While the B component is an M2 V companion at a distance of
$\approx$100 au (Krisciunas et al. 1993), the orbit of that companion is not
well known enough to determine the mass function of the system. Adopting a mass
range of 1.5--1.7\,$M_{\odot}$, which is reasonable for F0V stars (Popper
1980), and further $R=1.64R_{\odot}$ (Mantegazza et al.\ 1994), we find $K_1\in
[40,45]$ for $f_1$ and $K_2\in [211,240]$ for $f_2$.  It is clear from these
numbers that we are dealing with high-order $g$-modes: the transverse
displacement is dominant over the radial one.

The equivalent width of the profiles varies by 8\% for $f_1$
and by 3\% for $f_2$.
Since equivalent width variations are usually interpreted in
terms of temperature variations, these findings are in agreement with the
photometric variations in the sense that the most important temperature
variation is due to $f_1$.

It is clear from Fig\,\ref{mom} that some of the theoretical amplitudes of
the moments are uncertain. For such data the discriminant defined in
equation\,(\ref{discr}) is much better than the one defined by Aerts et
al.\ (1992), because the former is constructed in such a way that the most
accurate amplitudes are given the largest weight.  The amplitudes of the
terms of the second moment varying with frequencies $f_1$ and $f_2$ are at
least as large as those varying with frequencies $2f_1$ and $2f_2$.  This
is a signature of the presence of non-axisymmetric modes (see Aerts et
al.\ 1992).

The amplitudes of the moments are used to calculate the discriminants
$\Gamma_{\ell}^m(v_{\rm p},i,\sigma)$ and their minima $\gamma_{\ell}^m$ for
each candidate mode $(\ell,m)$. We have used the upper limits for the
$K$-values as input numbers for the discriminant. In the case of V398\,Aur, the
solution is independent of this choice since the discriminant searches the most
likely amplitude of the transverse velocity $Kv_{\rm p}$: a lower input $K$
gives a higher $v_{\rm p}$-value and vice versa. The adopted $K$-value is
important when the radial and transverse velocities are of the same order of
magnitude.  The results of the discriminant for $v\sin i=17.8$\,km s$^{-1}$ are
listed in Table\,\ref{gamma}.  We only give the best solutions in parameter
space. It is seen from the table that the two frequencies give rise to an
almost identical list of most likely modes. This situation is comparable to the
results found for the $\beta\,$Cephei star $\beta\,$CMa (Aerts et al.\ 1994)
and are interpreted in terms of the presence of two modes with identical degree
$\ell$ but opposite azimuthal numbers $m$. The combination of two $\ell=3$
modes, one with $m=+1$ and one with $m=-1$, seems most likely here, since the
inclination and intrinsic width fits nicely for such a combination, while this
is less the case for the $\ell=4$ or 2 solutions. Nevertheless, we cannot
exclude the possibility of other combinations of modes given in
Table\,\ref{gamma}, since the error on the inclination can be quite large. It
does seem fair to conclude that $|m|=1$ for both modes.  We have recalculated
the discriminant for the lower limit of the projected rotation velocity, i.e.\
$v\sin i=15\,$km s$^{-1}$. The results are comparable to the ones given in
Table\,\ref{gamma}, but with slightly different values of the inclination, the
amplitude, and the intrinsic width.

The question then arises if the two modes can be due to rotational splitting.
Dziembowski \& Goode (1992, equations 22 and 117) give an expression for the
observed frequencies $f$ expected from rotational splitting of a mode
$(\ell,m)$ in the case of rigid rotation and in the limit of high-order
$g$-modes~:
\begin{equation}
\renewcommand{\arraystretch}{2.5}
\begin{array}{lll}
f&=&\displaystyle{f_0-m\Omega\left(1-\frac{1}{\ell(\ell+1)}\right)}\\ &&
\displaystyle{-\frac{m^2\Omega^2}{f_0}
\frac{4\ell(\ell+1)\left(2\ell(\ell+1)-3\right)-9}
{2\ell^2(\ell+1)^2\left(4\ell(\ell+1)-3\right)}},
\end{array}
\label{split}
\end{equation}
where $f_0$ is the frequency in the corotating frame in the case of the
absence of rotation. This formula is derived under the assumption that
$\Omega<\!\!<f_0$.  Uniform spacing occurs if the second-order term in
$\Omega$ is negligible.  If we assume that this is the case, then
$f_0=0.571\,$d$^{-1}$.  If the $\ell=3, m=+1$ solution belongs to $f_2$
and the $\ell=3, m=-1$ one to $f_1$, then we obtain this $f_0$ from $f_2$
and from $f_1$ for $\Omega=0.245\,$d$^{-1}$.  Assuming again that $v\sin
i=17.8\,$km s$^{-1}$ and a stellar radius of $1.64R_{\odot}$ (Mantegazza
et al.\ 1994), such a rotation frequency corresponds to an inclination
angle of $61^{\circ}$, a value fully in agreement with the ones derived
from the two discriminants. For $f_0=0.571\,$d$^{-1}$, the second-order
term in equation\,(\ref{split}) amounts to only 0.008\,d$^{-1}$.  None of
the other combinations of $\ell_1=\ell_2, m_1=-m_2$ listed in
Table\,\ref{gamma} have an inclination that is close to $61^{\circ}$ for
both modes.  We conclude that the two modes of V398\,Aur can be due to
rotational splitting of an $\ell=3$ mode.  Unfortunately, the suggestion
of Krisciunas et al.\ (1995) that the ``true'' projected rotational
velocity may be less than the observed mean value cannot be confirmed by a
comparison of the results of the discriminants for different $v\sin
i$-values, because the uncertainty on the inclination is unknown. The true
ratios of the horizontal to vertical velocity amplitudes for uniform
splitting of the $\ell=3$ mode are $K_1\in [83,94], K_2\in [72,82]$.
There is no evidence of the presence of other frequency components of the
uniform rotational splitting belonging to $\ell=3$ in the photometry.

Assuming a rotation frequency $\Omega$ of 0.245\,d$^{-1}$ and a frequency
$f_0=0.571\,$d$^{-1}$ as found in the case of uniform rotational splitting of
the two $\ell=3$ modes, we obtain $\Omega/f_0=43\%$.  With such a high ratio,
the frequency-splitting formula given in equation\,(\ref{split}) may not be
very accurate, but it is the best one available. Also, the velocity field of an
NRP can no longer be described in terms of one $(\ell,m)$-value, because the
Coriolis force and the centrifugal forces are not negligible.  The Coriolis
force induces toroidal correction terms described by an $Y_{\ell+1}^m$ and an
$Y_{\ell-1}^m$ spherical harmonic in the case of an NRP with wavenumbers
$(\ell,m)$ in the non-rotating case (see e.g.\ Aerts \& Waelkens 1993).  It is
clear that our results found from the moment method, which does not take into
account these corrections, have to be interpreted with caution. It is possible
that some of the solutions with $|m|=1$ found by the discriminant are
manifestations of the toroidal corrections since the latter have the same
azimuthal number as the non-rotation spherical harmonic. If this is correct,
then it would be, as far as we know, the first detection of toroidal
corrections in a real star.

In order to have an idea of the line-profile variations (LPVs) expected
for the combination of the two $\ell=3$ modes as found by the
discriminant, we have generated theoretical LPVs with the code presented
by Aerts \& Waelkens (1993). This code takes into account the toroidal
correction terms due to the Coriolis force by means of a perturbation
analysis assuming that $\Omega/f_0<\!\!<1$, but not those related to the
centrifugal forces. As far as we know, no line-profile code is available
that does take into account the latter effects. Because of the large
$\Omega/f_0$-value, it is questionable if the perturbation analysis is
accurate. With this in mind, we do not expect perfect fits, the more so
since our code does not include equivalent width variations and assumes an
intrinsic Gaussian that is time-independent. It is clear that these
conditions are not exactly fulfilled.  Furthermore, the effective temperature
of the hemisphere of the star facing us is not constant.  From the variations
of $B$--$V$ color of the star (Table 3 and Fig 6 of Krisciunas et al. 1995)
and the rate  of change of temperature of mid-main sequence stars as a
function of $B$--$V$ (Allen 1973), we estimate for V\,398 Aur
that $T_{\rm eff}$
varies by $\approx$ 100\,K.  Nevertheless, we can be confident that the
solution proposed by the discriminant is good if the obtained theoretical
LPVs are compatible with the observed ones.

The width of the intrinsic profile found by the discriminant must be an
overestimate of the true value, since the velocity contributions of the
toroidal terms were neglected in the non-rotating model. This width can,
however, easily be found from the observed LPVs once a mode identification
has been obtained. We have found that $\sigma=5\,$km s$^{-1}$ is a more
realistic value.  Also the pulsation amplitudes will be overestimated by
the discriminant for the same reason. On the other hand, the $K$-values
used were also overestimated.  We have generated theoretical
LPVs with all the parameters except the $v_{\rm p}$-values fixed (we have
again chosen the upper limit for both $K$-values of zeroth-order in $\Omega$,
but as explained above,
this choice does not matter). In this way, a better estimate of the
$v_{\rm p}$-values could be obtained for the two modes.  Finally, the
phase shift between the two modes was determined from the first moment.
The results of our theoretically generated LPVs for the best choice of the
pulsation amplitudes are shown in Fig\,\ref{fits}, where we compare them
with some of the observed profiles. We show 10 arbitrarily chosen observed
profiles that are spread in phase according to $f_2$ with respect to an
arbitrary reference epoch.  The solid lines are the profiles obtained with
the code that takes into account the toroidal terms ($\Omega/f_0=0.43$),
while the dashed profiles are those obtained with a model that neglects
the rotation effect (i.e.\ those with $\Omega/f_0=0.0$).  As other input
parameters, we used~:
\renewcommand{\arraystretch}{1.2}
\begin{equation}
\label{opl}
\left\{\begin{array}{l}
f_1=0.795\mbox{d}^{-1}, \ell_1=3, m_1=-1, \\
\ \ v_{\rm p}^1=0.100\,\mbox{km s}^{-1}, K_1=94,
     K_1v_{\rm p}^1=9.4\,\mbox{km s}^{-1}\\
f_2=0.346\mbox{d}^{-1}, \ell_2=3, m_2=+1, \\
\ \ v_{\rm p}^2=0.025\,\mbox{km s}^{-1}, K_2=82,
      K_2v_{\rm p}^2=2.1\,\mbox{km s}^{-1}\\
v\sin i=17.8\,\mbox{km s}^{-1},\sigma=5\,\mbox{km s}^{-1},i=60^{\circ}.
\end{array}\right.
\end{equation}
\begin{figure}
\vspace{10truecm}
\caption[]{A comparison between observed and theoretical LPVs for the
parameters listed in (\ref{opl}).  The dots are actual observed points
from the cross-correlation profiles obtained by R. F. Griffin.  The solid
lines are profiles based on a theoretical model taking toroidal terms into
account.  The dashed lines are profiles for a theoretical model that
neglects rotational effects.}
\label{fits}
\end{figure}
First of all, the observed and calculated profiles show the same global
behavior such that our proposed solution of the velocity parameters listed in
(\ref{opl}) remains valid. The variation in line depth is accounted for by some
of the theoretical profiles, but not by all of them.  The evolution of the two
sets of theoretical profiles during the cycle are not too different from each
other. It seems that the profiles without the toroidal terms are better during
the first part of the cycle, while the others are better during the second
part. The effect of the toroidal terms on LPVs depends completely on the kind
of modes. It was studied by Aerts \& Waelkens (1993) for some monoperiodic
$p$-modes but has not yet been studied in the case of $g$-modes or
multiperiodicity. A detailed study about this is currently being undertaken
(Schrijvers et al., in preparation).  The fact that the differences between the
two sets of profiles are not very large makes us confident that the results of
the discriminant are not too bad.  We have also calculated theoretical profiles
for the other candidate rotationally splitted modes listed in
Table\,\ref{gamma}, i.e. the $\ell=4,|m|=1$ and the $\ell=2,|m|=1$ solution. In
doing so, we have taken the average of the two listed $i$-values. In both
cases, the correspondence between the observed profiles and the theoretical
ones is worse compared to the $\ell=3,|m|=1$ case described by (\ref{opl}).

\section{Discussion}

Zerbi et al. (1996) have analyzed the photometric variations of V398\,Aur
using data from the 1994/5 observing season.  While they find essentially
the same principal frequency in the power spectrum of the photometric data
($f_{1}$ = 0.796 d$^{-1}$), they do not find a {\em single} peak at
$f_{2}$ = 0.346 d$^{-1}$. Instead they find multiple frequencies in the
range $0.28 < f < 0.49$ d$^{-1}$.  It is known that the {\em amplitudes}
of the sinusoids making up the photometric variations are {\em not constant}
(Krisciunas et al. 1995); these authors make a case that the {\em
frequencies} $f_{1}$ and $f_{2}$ were constant from 1992 January to 1994
February.  The large data
set obtained by Zerbi et al. in 1994/5 allows us to wonder if the
frequencies in the power spectrum of the photometry of V398\,Aur are also
variable. In any case, it was found from the spectroscopic data of the
1993/4 season that the frequency 0.346\,d$^{-1}$ is the most pronounced
one in all three moment variations.

It was emphasized by Krisciunas et al.\ (1995) that the difference in
photometric and spectroscopic behavior of the frequencies $f_1$ and $f_2$
can be understood if $f_2$ would be due to a low-degree mode and $f_1$ to
a higher-degree one. This explanation is no longer appropriate if our
identification is correct, since both modes have the same geometrical
representation. From the parameters given in (\ref{opl}), we note that the
amplitude of the radial part of the pulsation for $f_1$ is a factor
4 larger than the one for $f_2$, while its angular dependence is the same.
Since the photometric variability is determined most of all by temperature
variations and since the latter are determined by the radial part of the
pulsation, it is quite understandable that the photometric variability is
dominated by the mode with frequency $f_1$.

For the spectroscopic behavior, the radial displacement can completely be
neglected in the case of large transverse amplitudes as we have here. The
relative importance of both modes for the line profiles could be the following.
The displacement field $\vec{\xi}$ correct up to first order in $\Omega/f_0$ in
the case of the two $\ell=3$ modes can easily be calculated, e.g.\ from the
paper by Aerts \& Waelkens (1993).  The toroidal corrections for the prograde
mode $(m=-1)$ and for the retrograde mode $(m=+1)$ are different.  Therefore,
it is possible that the total transverse velocity is larger for the mode with
frequency $f_2$ than for the mode with frequency $f_1$, despite its smaller
spheroidal amplitude $KA$. Quite understandably, the line profiles are then
mostly determined by $f_2$.

In order to test this point of view, we have explicitly calculated the
velocities in the direction of the observer for both modes, taking into account
the toroidal corrections. We indeed find that, although the retrograde mode has
a four times smaller spheroidal zero-rotation amplitude, its total velocity
component in the line of sight becomes larger due to the toroidal correction
terms. The maximum velocity reached by the retrograde mode is a factor 1.8
larger than the one of the prograde mode.

It thus seems possible that the toroidal corrections that appear due to
influence of the rotation on a spheroidal mode introduce a different
photometric and spectroscopic behavior when both prograde and retrograde modes
are involved.  To our knowledge, this is the first detected consequence of the
presence of such toroidal corrections. Our point of view could be further
verified by means of an analysis of a large set of high-resolution, high S/N
spectra of a star pulsating in both prograde and retrograde modes and having a
large rotation frequency, i.e.\ having $\Omega/f_0>20\%$.

\section*{ACKNOWLEDGMENTS}

We are grateful to Dr. M. Mayor and the Observatoire de $\rm Gen\grave{e}ve$
for having made the 1-m Geneva telescope and Coravel radial velocity
spectrometer available to Dr. R. F. Griffin for the data analyzed here.
We thank Dr. Griffin for allowing himself to be convinced that the radial
velocity of V\,398 Aur might indeed be variable and worth looking into by
means of systematic observations.

\bsp

\end{document}